\documentstyle[aps,twocolumn]{revtex}

\tighten
\draft

\begin{document}
\title{Kondo spin liquid and magnetically long-range ordered states in the Kondo
necklace model }
\author{Guang-Ming Zhang$^1$, Qiang Gu$^1$, and Lu Yu$^{2,3}$}
\address{$^1$Center for Advanced Study, Tsinghua University, Beijing 100084, P. R.
China.\\
$^2$International Center for Theoretical Physics, P. O. Box 586, Trieste
34100, Italy.\\
$^3$ Institute for Theoretical Physics, Academic Sinica, Beijing 100080, P.
R. China}
\date{\today}
\maketitle

\begin{abstract}
A simplified version of the symmetric Kondo lattice model, the Kondo
necklace model, is studied by using a representation of impurity and
conduction electron spins in terms of local Kondo singlet and triplet
operators. Within a mean field theory, a spin gap always appears in the spin
triplet excitation spectrum in 1D , leading to a Kondo spin liquid state for
any finite values of coupling strength $t/J$ (with $t$ as hopping and $J$ as
exchange); in 2D and 3D cubic lattices the spin gaps are found to vanish
continuously around $(t/J)_c\approx 0.70$ and $(t/J)_c\approx 0.38$,
respectively, where quantum phase transitions occur and the Kondo spin
liquid state changes into an antiferromagnetically long-range ordered state.
These results are in agreement with variational Monte Carlo, higher-order
series expansion, and recent quantum Monte Carlo calculations for the
symmetric Kondo lattice model. \newline
{PACS numberes: 71.27.+a, 71.10.Hf, 75.10.Jm, 75.30.Hx}\newline
\end{abstract}

\smallskip

Since the discovery of the class of stoichiometric insulating compounds
known as Kondo insulators \cite{aeppli}, there has been revived interest in
the symmetric Kondo lattice Hamiltonian, 
\begin{equation}
H=-t\sum_{{\bf \langle }i,j{\bf \rangle }}(C_{i,\sigma }^{\dagger
}C_{j,\sigma }+h.c.)+J\sum_i{\bf S}_i\cdot C_{i,\alpha }^{\dagger }{\bf %
\sigma }_{\alpha \beta }C_{i,\beta },  \nonumber
\end{equation}
as a model of concentrated magnetic impurity spins coupled to conduction
electrons. One of the important issues is the interplay between the Kondo
screening and the magnetic interactions among localized spins mediated by
the conduction electrons. The former effect favors a nonmagnetic Kondo spin
liquid (singlet) state, while the latter interactions tend to stabilize an
antiferromagnetically (AF) long-range ordered state. The character of such a
transition between these two distinct phases has been a long standing issue
since it was first pointed out by Doniach \cite{doniach}. There have been a
lot of investigations for the symmetric one-dimensional model, showing that
its ground state is a disordered Kondo spin liquid state for {\it any}
finite values of the coupling strength $t/J$ \cite{review}. For two and
three dimensional models, however, various approximate approaches such as
variational Monte Carlo calculations \cite{wang}, higher-order series
expansions \cite{shi}, quantum Monte Carlo simulations \cite{assaad,vekic},
and infinite dimensional calculations \cite{jarrell}, suggest that the Kondo
spin liquid state may change into an AF long-range ordered state at certain
value of the coupling strength at low temperatures.

Since there are a lot of difficulties in directly attacking the symmetric
Kondo lattice model even in the 1D case, a simplified version called Kondo
necklace model was introduced by Doniach \cite{doniach}, 
\begin{equation}
H=t\sum_{\langle i,j\rangle }(\tau _i^x\tau _j^x+\tau _i^y\tau _j^y)+J\sum_i%
{\bf S}_i\cdot {\bf \tau }_i,
\end{equation}
where both ${\bf \tau }_i$ and ${\bf S}_i$ are spin 1/2 Pauli operators,
denoting the conduction electron spin and impurity spin operators,
respectively, and $\langle i,j\rangle $ means summation over the nearest
neighbor conduction electron sites. Actually this simplified model is
meaningful in general D dimensions ($D=1,2,3$) in its own right. Due to the
suppression of charge fluctuations in the symmetric model, the charge
degrees of freedom are frozen out, so the first term of Eq.(2) represents
the spin degrees of freedom imitating the propagation of the conduction
electrons. This can be clearly seen in the 1D case, where the first term is
equivalent after a Jordan-Wigner transformation to a band of spinless
fermions, which interact with localized spins via an AF spin-spin exchange
coupling \cite{doniach}. 

Although the simplified model has only U(1) spin symmetry, lower than SU(2)
for the Kondo lattice model, the essential feature of these two models is
kept. Thus, one would expect that the main physical properties of the
original symmetric Kondo lattice model should be maintained in the Kondo
necklace model. However, most of approaches used to treat the 1D Kondo
necklace model, including the variational mean-field calculation \cite
{doniach}, approximate real-space renormalization group theory \cite{jullien}%
, and recent finite size scaling analysis \cite{solyom}, have found a finite
critical value of coupling strength $(J/t)_c\sim 0.24-0.38$, below which an
AF quasi-long-range order state appears, {\it in contrast to} $J_c=0$, the
result of quantum Monte Carlo simulation for the 1D Kondo necklace model 
\cite{scalettar} and the numerical result for the 1D symmetric Kondo lattice
model \cite{pfeuty,tsunetsugu}. It is thus controversial whether the
simplified spin model can be used to approximate the original symmetric
Kondo lattice model. In this paper, we try to resolve this issue, starting
from the Kondo necklace model, using the Kondo spin singlet and triplet
representation, to reproduce correct ground states of the symmetric Kondo
lattice model. In the 1D case, the system is found to be in a Kondo spin
liquid state with a finite spin gap for any finite $t/J$, while on 2D and 3D
cubic lattices a quantum phase transition occurs around $(t/J)_c\sim 0.70$
and $(t/J)_c\sim 0.38$, respectively, where the Kondo spin liquid state
changes into an AF long-range ordered state, in excellent agreement with the
variational Monte Carlo calculation \cite{wang}, higher-order series
expansion \cite{shi}, and recent quantum Monte Carlo simulation \cite{assaad}%
, on the corresponding symmetric Kondo lattice model.

Our starting point is the strong coupling limit $t=0$, where the lowest
energy state of the model Hamiltonian Eq.(2) reduces to a sum over
contributions from independent local Kondo spin singlet states at each
lattice site. When $t\neq 0$, interactions between these independent local
Kondo spin singlets are switched on. It will be seen later that this leads
to very reasonable results even for $t\geq J$, which is of interest here.
Usually, for two $s=1/2$ spins ${\bf \tau }_i$ and ${\bf S}_i$ placed on a
lattice site, the local Hilbert space is spanned by four states consisting
of one singlet and three triplet states defined as being created out of the
vacuum $|0\rangle $ by the singlet and triplet creation operators: $%
|s\rangle =s^{\dagger }|0\rangle $ and $|t_\alpha \rangle $ $=t_\alpha
^{\dagger }|0\rangle $, ($\alpha =x,$ $y,$ $z$). A representation of the
impurity spins and conduction electron spins in terms of these singlet and
triplet operators is given by

\begin{eqnarray}
S_{n,\alpha } &=&\frac 12(s_n^{\dagger }t_{n,\alpha }+t_{n,\alpha }^{\dagger
}s_n-i\epsilon _{\alpha \beta \gamma }t_{n,\beta }^{\dagger }t_{n,\gamma }),
\nonumber \\
\tau _{n,\alpha } &=&\frac 12(-s_n^{\dagger }t_{n,\alpha }-t_{n,\alpha
}^{\dagger }s_n-i\epsilon _{\alpha \beta \gamma }t_{n,\beta }^{\dagger
}t_{n,\gamma }),
\end{eqnarray}
where $\alpha $, $\beta $, and $\gamma $ represent components along the $x$, 
$y$, and $z$ axes, respectively, and $\epsilon $ is the antisymmetric
Levi-Civita tensor. This type of spin representation in terms of singlet and
triplet (bond) operators was first proposed by Sachdev and Bhatt to study
the properties of dimerized phases \cite{sachdev} and then it was
successfully used to consider the spin ladders\cite{rice} and $s=1$
antiferromagnetic Heisenberg spin chains\cite{hanting}. As shown later, this
representation faithfully describes the low temperature physics in the
symmetric Kondo lattice model. In order to restrict the physical states to
either singlets or triplets, a local constraint is introduced: $s_n^{\dagger
}s_n+\sum\limits_\alpha t_{n,\alpha }^{\dagger }t_{n,\alpha }=1.$ Taking the
singlet and triple operators at each site to satisfy bosonic commutation
relations: $[s_n,s_n^{\dagger }]=1$, $[t_{n,\alpha },t_{n,\beta }^{\dagger
}]=\delta _{\alpha ,\beta }$, and $[s_n,t_{n,\alpha }^{\dagger }]=0$, the
SU(2) algebra of the spins ${\bf \tau }_n$ and ${\bf S}_n$ can be reproduced,

\begin{eqnarray}
\lbrack S_{n,\alpha },S_{n,\beta }] &=&i\epsilon _{\alpha \beta \gamma
}S_{n,\gamma },\text{ \quad }[\tau _{n,\alpha },\tau _{n,\beta }]=i\epsilon
_{\alpha \beta \gamma }\tau _{n,\gamma },  \nonumber \\
\lbrack S_{n,\alpha },\tau _{n,\beta }] &=&0,\text{ \quad }{\bf S}_n^2={\bf %
\tau }_n^2=\frac 34.
\end{eqnarray}
Substituting the operator representation of the impurity and conduction
electron spins, we obtain the following form of the model Hamiltonian,

\begin{eqnarray}
H &=&H_0+H_1+H_2+H_3,  \nonumber \\
H_0 &=&\frac J4\sum_i(-3s_i^{\dagger }s_i+\sum_\alpha t_{i,\alpha }^{\dagger
}t_{i,\alpha })  \nonumber \\
&&\text{ \quad }+\sum_i\mu _i(s_i^{\dagger }s_i+\sum_\alpha t_{i,\alpha
}^{\dagger }t_{i,\alpha }-1),  \nonumber \\
H_1 &=&\frac t4\sum_{\langle ij\rangle }\left[ s_i^{\dagger }s_j^{\dagger
}\left( t_{i,x}t_{j,x}+t_{i,y}t_{j,y}\right) \right.  \nonumber \\
&&\text{ \quad }+\left. s_i^{\dagger }s_j\left( t_{i,x}t_{j,x}^{\dagger
}+t_{i,y}t_{j,y}^{\dagger }\right) +h.c.\right] ,  \nonumber \\
H_2 &=&-\frac t4\sum_{\langle ij\rangle }\left[ t_{i,z}^{\dagger
}t_{j,z}^{\dagger }\left( t_{i,x}t_{j,x}+t_{i,y}t_{j,y}\right) \right. 
\nonumber \\
&&\text{ \quad }-\left. t_{i,z}^{\dagger }t_{j,z}\left(
t_{i,x}t_{j,x}^{\dagger }+t_{i,y}t_{j,y}^{\dagger }\right) +h.c.\right] , 
\nonumber \\
H_3 &=&\frac{it}4\sum_{\langle ij\rangle }\sum_{\alpha ,\beta ,\gamma
}\epsilon _{\alpha \beta \gamma \text{ }}\left[ s_i^{\dagger }t_{i,\alpha
}t_{j,\beta }^{\dagger }t_{j,\gamma }+s_j^{\dagger }t_{j,\alpha }t_{i,\beta
}^{\dagger }t_{i,\gamma }+h.c.\right],
\end{eqnarray}
where a site-dependent chemical potential $\mu _i$ has been introduced to
impose the local constraint. Here the local spin triplet states are split
into two parallel spin states with $m_s=\pm 1$ and an anti-parallel spin
state with $m_s=0$. $H_1$ describes the couplings between the singlet state
and the parallel spin triplet states, while $H_2$ corresponds to the
couplings of the parallel spin and the anti-parallel spin triplet states. $%
H_3$ describes an interaction of one singlet boson and three different
components of triplet bosons.

The above Hamiltonian can be solved by a mean field decoupling of the
quartic terms. It yields an effective Hamiltonian $H_{mf}$ with only
quadratic operators. We take $\langle s_i^{\dagger }\rangle =\langle
s_i\rangle =\overline{s}$, which corresponds to {\it a condensation of the
local Kondo spin singlets on each site }in accordance with the configuration
of the ground state in the strong coupling limit, and the local chemical
potential is replaced by a global one. We will consider here only the terms $%
H_0$ and $H_1$, as it can be shown that inclusion of $H_2$ changes the
results only slightly \cite{rice,gmzhang} and all the decouplings of $H_3$
identically vanish within the present mean field theory. After performing a
Fourier transformation of the boson operators, $t_{i,\alpha }=\frac 1{\sqrt{N%
}}\sum\limits_{{\bf k}}t_{{\bf k},\alpha }$ $e^{-i{\bf k\cdot r}_i}$, the
mean field effective Hamiltonian is given by 
\begin{eqnarray}
&& H_{mf}=N\left( -\frac 34J\text{ }\overline{s}^2+\mu \overline{s}^2-\mu
\right) +\left( \frac J4+\mu \right) \sum_{{\bf k}}t_{{\bf k},z}^{\dagger
}t_{{\bf k},z}  \nonumber \\
&& \hspace{0.5cm} +\sum_{{\bf k},\beta =x,y}\left[ \Lambda _{{\bf k}}t_{{\bf %
k},\beta }^{\dagger }t_{{\bf k},\beta }+\Delta _{{\bf k}}\left( t_{{\bf k}%
,\beta }^{\dagger }t_{-{\bf k},\beta }^{\dagger }+t_{{\bf k},\beta }t_{-{\bf %
k},\beta }\right) \right] ,
\end{eqnarray}
with $\Lambda _{{\bf k}}=\left( \frac J4+\mu \right) +\frac 12t\overline{s}%
^2\lambda ({\bf k)}$, $\Delta _{{\bf k}}=\frac 14t\overline{s}^2\lambda (%
{\bf k)}$, and $\lambda ({\bf k)=}\sum\limits_{a=1}^d\cos k_a$. The lattice
spacing has been taken to be unity. This mean field Hamiltonian can be
diagonalized by a Bogoliubov transformation into new boson operators: $%
\widetilde{t}_{{\bf k},\beta }=u_{{\bf k}}t_{{\bf k},\beta }+v_{{\bf k}}t_{-%
{\bf k},\beta }^{\dagger }$, where the coefficients $u_{{\bf k}}$ and $v_{%
{\bf k}}$ are even functions of ${\bf k}$, and are determined to be: $u_{%
{\bf k}}^2+v_{{\bf k}}^2=\cosh 2\theta _{{\bf k}}=\frac{\Lambda _{{\bf k}}}{%
\sqrt{\Lambda _{{\bf k}}^2-(2\Delta _{{\bf k}})^2}}$ and $2u_{{\bf k}}v_{%
{\bf k}}=\sinh 2\theta _{{\bf k}}=-\frac{2\Delta _{{\bf k}}}{\sqrt{\Lambda _{%
{\bf k}}^2-(2\Delta _{{\bf k}})^2}}.$ Then we obtain 
\begin{equation}
H_{mf}=\omega _0\sum_{{\bf k}}t_{{\bf k},z}^{\dagger }t_{{\bf k},z}+\sum_{%
{\bf k},\beta =x,y}\omega _{{\bf k}}\widetilde{t}_{{\bf k},\beta }^{\dagger }%
\widetilde{t}_{{\bf k},\beta }+E_g,
\end{equation}
where $\omega _0=\left( \frac J4+\mu \right) $ is the dispersionless energy
level of the anti-parallel spin triplet excited state, $\omega _{{\bf k}}=%
\sqrt{\Lambda _{{\bf k}}^2-(2\Delta _{{\bf k}})^2}$ corresponds to the
excitation spectrum of the parallel spin triplet excited states, and the
ground state energy of the system is $E_g=N\left( -\frac 34J\text{ }%
\overline{s}^2+\mu \overline{s}^2-\mu \right) +\sum\limits_{{\bf k}}\left(
\omega _{{\bf k}}-\Lambda _{{\bf k}}\right) $. By minimizing the ground
state energy with respect to $\mu $ and $\overline{s}$, we derive the
following saddle-point equations,

\begin{eqnarray}
\frac 1N\sum_{{\bf k}}\frac{\Lambda _{{\bf k}}}{\sqrt{\Lambda _{{\bf k}%
}^2-(2\Delta _{{\bf k}})^2}} &=&(2-\overline{s}^2),  \nonumber \\
\frac tN\sum_{{\bf k}}\sqrt{\frac{\Lambda _{{\bf k}}-2\Delta _{{\bf k}}}{%
\Lambda _{{\bf k}}+2\Delta _{{\bf k}}}}\lambda ({\bf k)} &=&2J\left( \frac 34%
-\frac \mu J\right) .
\end{eqnarray}
When a dimensionless parameter $d=\frac tJ\frac{\overline{s}^2}{(\frac 14+%
\frac uJ)}$ is introduced, a self-consistent equation for $d$ can be obtained

\begin{equation}
d=\frac{2t}J\left[ 1-\frac 1{2N}\sum_{{\bf k}}\frac 1{\sqrt{1+d\lambda ({\bf %
k)}}}\right] ,
\end{equation}
to determine the variational parameters $\overline{s}$ and $\mu $ and the
spin triplet excitation spectra: $\omega _0=J\left( \frac 14+\frac \mu J%
\right) $ and $\omega _k=J\left( \frac 14+\frac \mu J\right) \sqrt{%
1+d\lambda ({\bf k)}}$. There is a minimum spin gap in the parallel spin
triplet spectrum at the AF reciprocal vector momentum ${\bf k}={\bf Q}$: $%
\Delta _{sp}=J\left( \frac 14+\frac \mu J\right) \sqrt{1-Zd/2}$, where $Z$
is the total number of the nearest neighbors on the cubic lattice.

In the 1D case, we first numerically calculate the parameters $d$, $%
\overline{s}^2$, and $\mu /J$ for a range of the coupling strength $0<t/J<$ $%
5$, and the minimum spin gap $\Delta _{sp}=J\left( \frac 14+\frac \mu J%
\right) \sqrt{1-d}$ is evaluated in the range of $0<t/J<$ $5$, which has
been delineated in Fig.1. The dispersive band can also be parameterized by a
spin density wave with a velocity given by $v_s=J\left( \frac 14+\frac \mu J%
\right) \sqrt{\frac d2}$. A linear drop of the spin gap is seen for small
values of $t/J$. As $t/J$ gets larger, the spin gap deviates considerably
from the linear behavior and there is no indication at all suggesting a
critical value for $t/J$ where the gap would vanish. Since the excitation
spectra are real and positive everywhere in the Brillouin zone, the system
will be in a quantum disordered --- Kondo spin liquid state for {\it finite}
values of the coupling strength $t/J$, and the spin-spin correlation
function decays exponentially at large distances with a correlation length $%
\xi =\frac{v_s}{\Delta _{sp}}$. This is indeed consistent with both quantum
Monte Carlo simulations for the 1D Kondo necklace \cite{scalletar} and
numerical results for the 1D symmetric Kondo lattice model \cite
{pfeuty,tsunetsugu}. Thus, starting from the limit $t/J=0$ with localized
Kondo spin singlets on each site, we see that any finite coupling strength
delocalizes the local Kondo singlets, reducing the magnitude of the gap but
not closing it completely.

Having secured the correct ground state for the 1D symmetric Kondo lattice
model, we now turn to two and three dimensional ''Kondo necklace'' models on
a cubic lattice. In 2D, the variational parameters $d$, $\overline{s}^2$,
and $\mu /J$ can also be calculated from the saddle-point equations. The
minimum spin gap appears in the parallel spin triplet excitation at ${\bf k}%
=(\pi ,\pi )$: $\Delta _{sp}=J\left( \frac 14+\frac \mu J\right) \sqrt{1-2d}$%
, displayed in Fig.2. The most important feature here is that as the
coupling parameter $t/J$ increases, the drop of the spin gap in the small
values of $t/J$ continues down to the point $(t/J)_c\approx 0.70$ where the
spin gap actually vanishes. The critical coupling $(t/J)_c\approx 0.70$
corresponds to a quantum critical point for a phase transition from the
quantum disordered Kondo spin liquid to a magnetically long-range ordered
state. Surprisingly, the location of the critical point for the 2D Kondo
necklace model is {\it precisely} the value obtained from the variational
Monte Carlo calculation \cite{wang}, the higher-order series expansion \cite
{shi}, and recent quantum Monte Carlo simulation \cite{assaad} for the 2D
symmetric Kondo lattice model. When a similar calculation is carried out in
the 3D Kondo necklace model, the minimum spin gap appears at ${\bf k}=(\pi
,\pi ,\pi )$ and $\Delta _{sp}=J\left( \frac 14+\frac \mu J\right) \sqrt{1-3d%
}$, shown in Fig.3. As $t/J$ grows, the spin gap decreases and exhibits a
critical value $(t/J)_c\approx 0.38$, where the spin gap disappears
completely, showing a quantum phase transition from the quantum disordered
Kondo spin liquid to a magnetically long-range ordered state as well. This
transition point is in the same range as the higher-order series expansion 
\cite{shi} for the 3D symmetric Kondo lattice model: $(t/J)_c\approx 0.50$.

Moreover, the present mean field theory can also be applied to the
magnetically long-range ordered phase in the 2D and 3D Kondo necklace
models. If we assume that, not only the local Kondo spin singlets ($s$
bosons) condenses, one of the local Kondo spin triplets ($t_{{\bf k},x}$
bosons) condenses as well on the AF reciprocal vector $t_{_{{\bf k},x}}=%
\sqrt{N}\overline{t}\delta _{{\bf k,Q}}+\eta _{{\bf k},x}$, corresponding to 
{\it fixing the orientation of the localized spins along }$x$-{\it direction}%
, it will lead to another mean field effective Hamiltonian

\begin{eqnarray}
&&H_{mf}^{\prime }=E_g^{\prime }+\omega _0\sum_{{\bf k}}t_{{\bf k}%
,z}^{\dagger }t_{{\bf k},z}+\sum\limits_{{\bf k}}\omega _{{\bf k}}(%
\widetilde{t}_{{\bf k},y}^{\dagger }\widetilde{t}_{{\bf k},y}+\widetilde{%
\eta }_{{\bf k},x}^{\dagger }\widetilde{\eta }_{{\bf k},x}),  \nonumber \\
&&E_g^{\prime }=N\left[ -\frac 34J\overline{s}^2+\mu \overline{s}^2-\mu
+\left( \frac J4+\mu -\frac 12Zt\overline{s}^2\right) \overline{t}^2\right] 
\nonumber \\
&&\text{ \quad \qquad }+\sum\limits_{{\bf k}}\left( \omega _{{\bf k}%
}-\Lambda _{{\bf k}}\right) ,
\end{eqnarray}
where $\omega _{{\bf k}}$ has the same form as in the Kondo spin liquid
phase, and $\widetilde{\eta }_{{\bf k},x}^{\dagger }$ and $\widetilde{\eta }%
_{{\bf k},x}$ are the transverse spin triplet excitation mode. When the
order parameter $\overline{t}$ is nonzero, the saddle point equation for $%
\overline{t}$ yields $\mu =\frac 12Zt\overline{s}^2-\frac J4$, which makes
the parallel spin triplet excitation spectrum {\it gapless}: $\omega _{{\bf k%
}}=\frac 12Zt\overline{s}^2\sqrt{1+2\lambda ({\bf k)}/Z}.$ The ground state
corresponds to a magnetically long-range ordering state with a maximum
momentum ${\bf q=Q}$, and the mean field $\overline{t}$ represents the AF
order parameter. It has been suggested that a very appealing physical
picture of forming AF long-range order in the Kondo necklace or the
symmetric Kondo lattice models: when $t/J$ is small, the conduction electron
spins are locked and the impurity spins are screened completely, and the
ground state is a product of the local Kondo spin singlets -- quantum
disordered phase \cite{review}. As $t/J$ becomes larger and larger, the
conduction electrons (the spin degrees of freedom) have more possibility to
propagate to the nearest neighbor sites, and the localized magnetic impurity
spins is only {\it partially } screened{\it \ (}$\overline{s}\neq 0${\it )},
then the remaining part of the magnetic impurities on different lattice
sites start to develop long-range correlations ($\overline{t}\neq 0$)
mediated by the conduction electron spins \cite{sachdev}. Such a
magnetically long-range ordered state might be related to the ground states
of the U-based heavy fermion compounds (URu$_2$Si$_2$ and UPt$_3$) with a
very small magnitude of induced staggered magnetic moments. In order to
determine the parameters $\overline{s}$ and $\overline{t}$, we minimize the
ground state energy, derive the saddle point equations, and finally obtain 
\begin{eqnarray}
\overline{s}^2 &=&1+\frac J{Zt}-\frac 1{2N}\sum_{{\bf k}}\sqrt{1+2\lambda (%
{\bf k)/}Z},  \nonumber \\
\overline{t}^2 &=&1-\frac J{Zt}-\frac 1{2N}\sum_{{\bf k}}\frac 1{\sqrt{%
1+2\lambda ({\bf k)/}Z}}.
\end{eqnarray}
The AF order parameter is defined by $m_s=\overline{s}$ $\overline{t}$,
leading to the following expressions:

\begin{eqnarray*}
m_s &=&\sqrt{(0.35712-\frac J{4t})(0.52095+\frac J{4t})},\text{ \quad \quad
for 2D;} \\
m_s &=&\sqrt{(0.44234-\frac J{6t})(0.51263+\frac J{6t})},\text{ \quad \quad
for 3D.}
\end{eqnarray*}
These results have also been displayed in Fig.2 and Fig.3, respectively. In
Fig.2, our results are also compared with the numerical results for the spin
gap and staggered moment of the magnetic impurity spins in the recent
quantum Monte Carlo simulation on the 2D symmetric Kondo lattice model at
zero temperature \cite{assaad}.

In summary, we have presented a mean field theory for the Kondo necklace
model in 1D, 2D and 3D and have obtained their correct ground states
corresponding to the respective Kondo lattice model. A long standing
controversial issue has been thus resolved regarding the relationship
between these two models. As far as the spin part of the ground state
properties is concerned, the Kondo necklace model can reproduce the correct
phase diagrams of the symmetric Kondo lattice model at zero temperature.

{\bf Figure Captions}

\smallskip

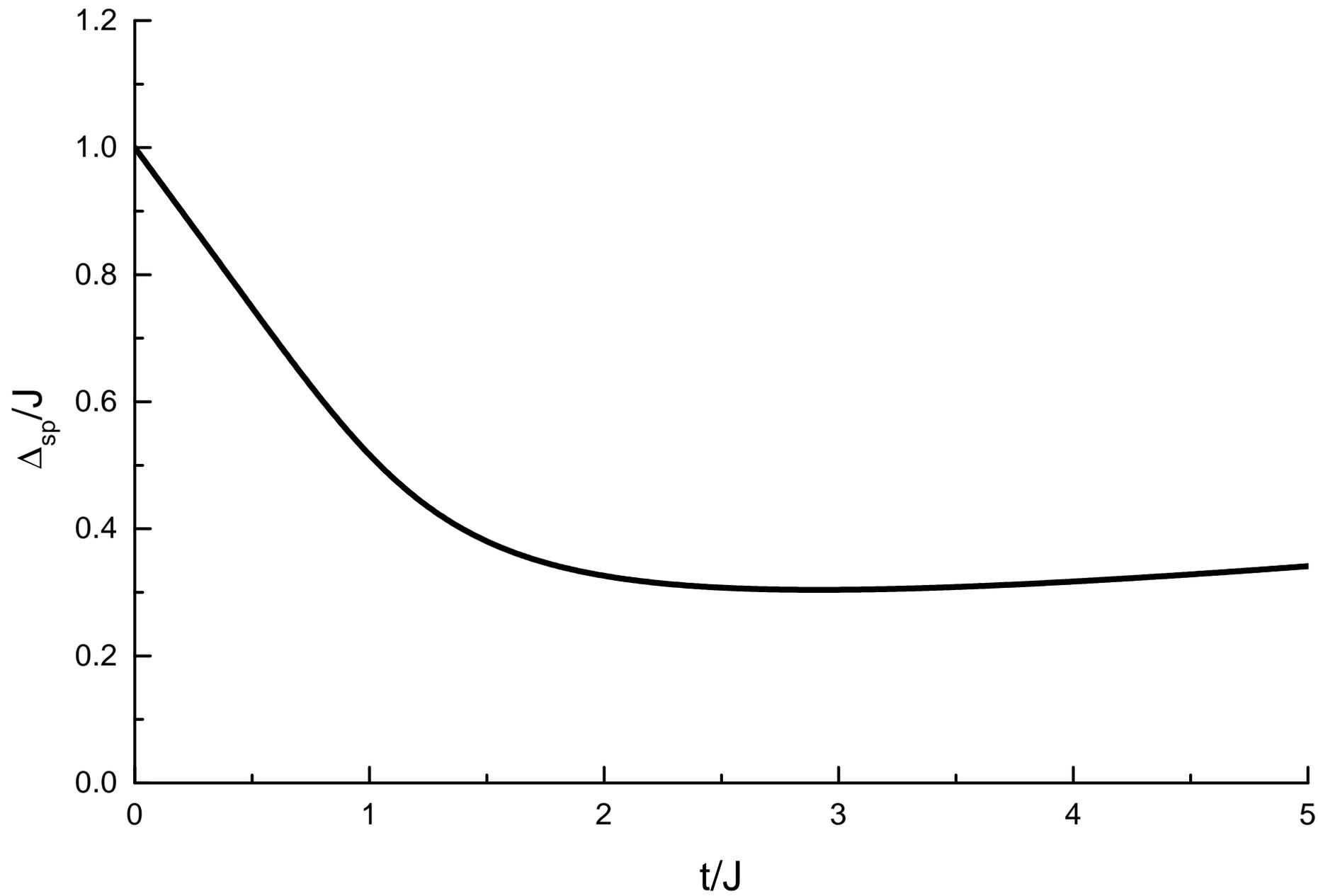
Fig.1. The variation of the spin gap upon increasing of the coupling
parameter $t/J$ of the 1D model at $T=0$.

\smallskip

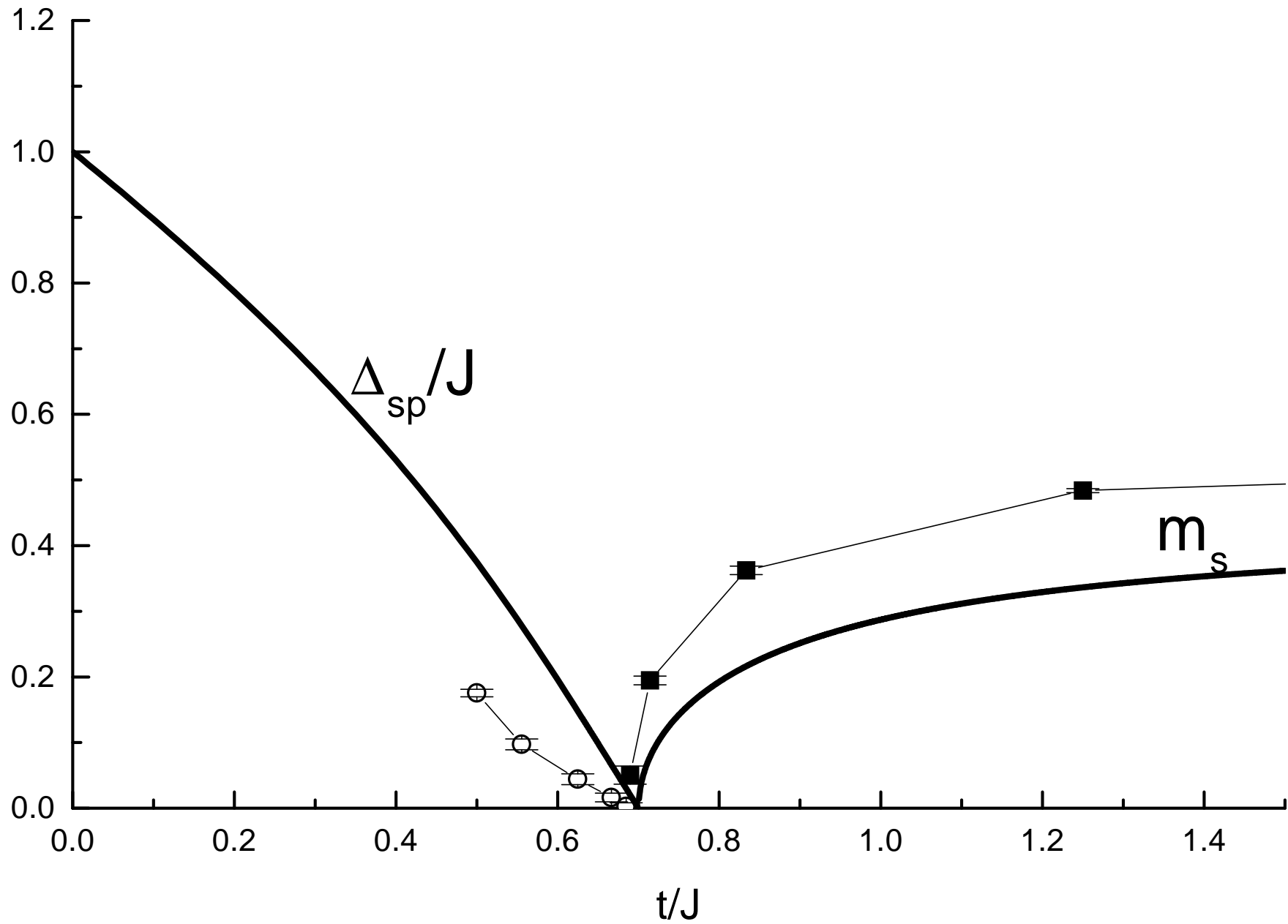
Fig.2. The spin gap and the staggered magnetic moment at zero temperature of
the 2D Kondo necklace model (bold line) in comparison with results of recent
quantum Monte Carlo simulation\cite{assaad} for the 2D Kondo lattice model.

\smallskip

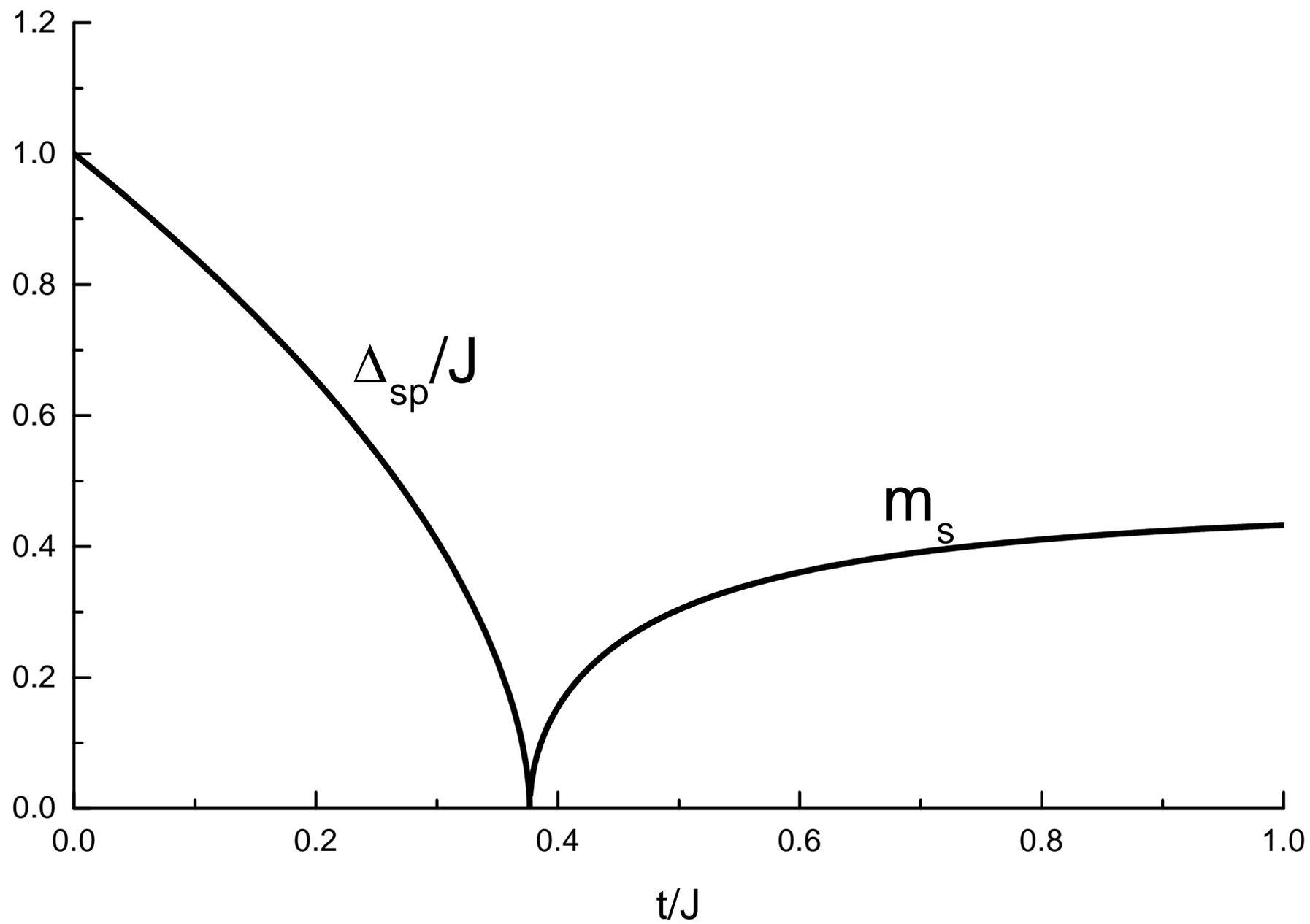
Fig.3. The spin gap and the staggered magnetic moment at zero temperature
for 3D Kondo necklace model.


\begin{references}
\bibitem{aeppli}  G. Aeppli and Z. Fisk, Comments on Condens. Matter Phys. 
{\bf 16}, 155 (1987).

\bibitem{doniach}  S. Doniach, Physica B {\bf 91}, 231 (1977).

\bibitem{review}  H. Tsunetsugu, M. Sigrist, and K. Ueda, Rev. Mod. Phys. 
{\bf 69}, 809 (1997), and references therein.

\bibitem{shi}  Z. P. Shi, R. R. Singh, M. P. Gelfand, and Z. Wang, Phys.
Rev. B {\bf 51}, 15630 (1995).

\bibitem{wang}  Z. Wang, X. P. Li, and D. H. Lee, Physica B {\bf 199-200},
463 (1984).

\bibitem{assaad}  F. F. Assaad, Phys. Rev. Lett. {\bf 83}, 796 (1999).

\bibitem{vekic}  M. Vekic, J. W. Cannon, D. J. Scalapino, and R. T.
Scalettar, and R. L. Sugar, Phys. Rev. Lett. {\bf 74}, 2367 (1995).

\bibitem{jarrell}  M. J. Rozenberg, Phys. Rev. B {\bf 52}, 7369 (1995).

\bibitem{jullien}  R. Jullien, J. N. Fields, and S. Doniach, Phys. Rev. B 
{\bf 16}, 4889 (1977); W. Hanke and J. E. Hirsch, {\it ibid}. {\bf 25}, 6748
(1982).

\bibitem{solyom}  P. Santini and J. Solyom, Phys. Rev. B {\bf 46}, 7422
(1992).

\bibitem{scalettar}  R. T. Scalettar, D. J. Scalapino. and R. J. Sugar,
Phys. Rev. B {\bf 31}, 7316 (1985).

\bibitem{pfeuty}  R. Jullien and P. Pfeuty, J. Phys. F 11, 353 (1981).

\bibitem{tsunetsugu}  H. Tsunetsugu, Y. Hatsugai, K. Ueda, and M. Sigrist,
Phys. Rev. B {\bf 46}, 3175 (1992); N. Shibata, T. Nishino, K. Ueda, and C.
Ishii,{\it \ ibid}. {\bf 53}, 8828 (1996).

\bibitem{sachdev}  S. Sachdev and R. N. Bhatt, Phys. Rev. B {\bf 41}, 9323
(1990).

\bibitem{rice}  S. Gopalan, T. M. Rice, and M. Sigrist, Phys. Rev. B {\bf 49}%
, 8901 (1994); B. Normand and T. M. Rice, {\it ibid}. {\bf 54}, 7180 (1996).

\bibitem{hanting}  Han-Ting Wang, Jue-Lian Shen, and Zhao-Bin Su, Phys. Rev.
B {\bf 56}, 14435 (1997).

\bibitem{gmzhang}  Guang-Ming Zhang, Qiang Gu, and Lu Yu, unpublished.
\end{references}
\end{document}